\documentstyle[epsfig,aps]{revtex}

\begin{document}
\title{Relaxation in Kinetic Models on Alternating Linear Chains}
\author{L. L. Gon\c {c}alves}
\address{Departamento de Fisica,\\
Universidade Federal do Cear\'{a},\\
Campus do Pici, C.P. 6030, 60451-970\\
Fortaleza, Cear\'{a}, Brazil}
\author{M. L\'{o}pez de Haro\cite{mariano}}
\address{Centro de Investigaci\'on en Energ\'{\i}a, UNAM, \\
Temixco, Morelos 62580, M\'{e}xico}
\author{J. Tag\"{u}e\~{n}a-Mart\'{\i}nez}
\address{Centro de Investigaci\'on en Energ\'{\i}a, UNAM, \\
Temixco, Morelos 62580, M\'{e}xico}
\date{\today}
\maketitle

\begin{abstract}
A restricted dynamics, previously introduced in a kinetic model for
relaxation phenomena in linear polymer chains, is used to study the dynamic
critical exponent of one-dimensional Ising models. Both the alternating
isotopic chain and the alternating-bond chain are considered. In contrast
with what occurs for the Glauber dynamics, in these two models the dynamic
critical exponent turns out to be the same. The alternating isotopic chain
with the restricted dynamics is shown to lead to Nagel scaling for
temperatures above some critical value. Further support is given relating
the Nagel scaling to the existence of multiple (simultaneous) relaxation
processes, the dynamics apparently not playing the most important role in
determining such scaling.
\end{abstract}




\section{Introduction.}

The scaling hypothesis of Halperin and Hohenberg \cite{Hohenberg} relates
the time scale $\tau $ and the correlation length $\xi $ and introduces the
dynamic critical exponent $z$. In the case of Ising models, $z$ was for a
long time believed to be universal, depending on the nature of conserved
quantities and of those features, for instance dimensionality \cite{Zhu},
which determine their static universality class. However, it is now well
established that for some simple systems this exponent is non universal\cite
{Kimball}-\cite{Robin}. In particular, in the case of one-dimensional
Glauber dynamics \cite{Glauber} the alternating isotopic chain\cite{Lind1}
presents universal behavior (in the sense that it leads to the same value of
the dynamic critical exponent as the homogeneous chain) whereas the
alternating-bond chain does not \cite{Lind1}-\cite{Tong} (see however Ref. 
\cite{Southern}). This breakdown of the dynamic scaling hypothesis is due to
the different energy barriers determining the diffusion constant and the
correlation length and hence entering into the value of the dynamic exponent 
$z$. This exponent is also typically strongly affected when the dynamics
takes place on a self-similar background. For instance, the study of the
Glauber dynamics in branching and non-branching Koch curves as well as in
the Sierpinsky gasket\cite{Achiam} indicates that $z$ is non universal.
Nevertheless, an analysis of the critical Glauber dynamics on the
Fibonacci-chain quasicrystal \cite{Robin} shows that the dynamic exponent $z$
is identical to that obtained for the alternating-bond Glauber chain.
Another interesting line of research that has also been followed \cite{Droz}
is to understand how different dynamics affect the universality. This work
is partly aimed as another contribution to such an understanding.

Kinetic Ising models have also been used in a variety of contexts not
limited to critical dynamics. In fact, in a previous paper \cite{LHTGE} we
introduced a quasi one-dimensional kinetic Ising-like model to study
relaxation phenomena in linear polymeric chains, including a region close to
the glass transition. In our original model the chains are made up of N
segments, each of which may be found in two possible orientations and the
Hamiltonian was chosen as to reduce to the one giving the intramolecular
energy of the Gibbs-di Marzio lattice model \cite{GdM}. For the stochastic
dynamics, we adopted a rule of transition for the configurational changes
which was tied to the creation or disappearance of flexes. As a consequence
of this restricted dynamics, only some states are selected, and in the
magnetic language, this implies that in the model the domain wall motion is
through a biased random walk. The purpose of this paper is, on the one hand,
to introduce the restricted dynamics into the two one-dimensional
alternating kinetic Ising models that have exact solutions with the Glauber
dynamics, namely the isotopic alternating chain and the alternating-bond
chain. On the other hand, we will use these extensions to examine the
dynamical critical exponent and, in the case of the isotopic chain, whether
the inclusion of several relaxation times may be related to multifractal
behavior. This last goal is motivated by the fact that recent dielectric
susceptibility measurements in several glass-forming systems and covering
wide ranges of temperature and frequency \cite{Nagel}-\cite{Wu}, have
suggested that the master curve in which all measurements are shown to scale
(Nagel plot) may be understood in terms of \ multiple relaxation processes.
In fact the existence of simultaneous relaxation processes is assumed to be
connected to multifractality much in the same way that this concept is
present in theories of chaos. Very recently we have shown that the
alternating isotopic chain with Glauber dynamics leads to Nagel scaling\cite
{PRL} and the question arises as to the role that a particular dynamics may
play in such a scaling. By examining the model with the restricted dynamics,
we aim at shedding some light into this issue.

The paper is organized as follows. Section 2 deals with the alternating
isotopic chain while section 3 is devoted to the alternating-bond chain. We
derive the wave-vector frequency dependent susceptibility, the equations
governing the time evolution of the wave-vector dependent magnetization and
the critical exponent $z$ in both cases. In section 4 we address the issue
of whether the restricted dynamics leads to the Debye-like scaling of the
response functions (as in the uniform chain) or whether the so-called Nagel
plot is useful in the case of the isotopic alternating chain. We close the
paper in section 5 with some concluding remarks.

\section{The alternating isotopic chain.}

The model consists of a closed linear chain with N sites occupied by two
isotopes (characterized by two different spin relaxation times) that are
alternately arranged. The Hamiltonian is the usual Ising Hamiltonian given by

\begin{equation}
H=-J\sum\limits_{j=1}^N\sigma _j\sigma _{j+1}  \label{1}
\end{equation}

\noindent where $\sigma _{j}$ is a stochastic (time-dependent) spin variable
assuming the values $\pm 1$ and $J$ the coupling constant. The configuration
of the chain is specified by the set of values $\left\{ \sigma _{1},\sigma
_{2},...\sigma _{N}\right\} $ at time $t$. As it will be argued below, the
Hamiltonian will turn out to be not all that relevant in the specific
calculations. Nevertheless it is necessary to define the states involved in
allowed transitions.

Instead of considering the Glauber dynamics, we assume a kind of transition
associated to the motion of domain walls. The idea is similar but not
identical to previous work by others in which either the domain wall motion
is strongly suppressed at low temperatures \cite{Kimball}\cite{Thol} or it
is through a one-dimensional random walk \cite{Skinner}. In our case the
transition associated with the $ith$ spin takes the form

\begin{eqnarray}
T_{i}\left\{ \sigma _{1},...,\sigma _{i},\sigma _{i+1},...,\sigma
_{N-1},\sigma _{N}\right\} \rightarrow  \nonumber \\
\left\{ \sigma _{1},...,\sigma _{i-1},-\sigma _{i},\sigma _{i+1},...,\sigma
_{N-1},\sigma _{N}\right\} ,  \label{2}
\end{eqnarray}

\noindent and we impose a biased random walk for the domain wall motion\cite
{Kaye}. In order to also account for the presence of the isotopes, we take
the transition probabilities to be given by

\begin{equation}
w_{i}(\sigma _{i-1},\sigma _{i})=\alpha _{i}(1-\gamma \sigma _{i-1}\sigma
_{i}).  \label{3b}
\end{equation}

Here, $\gamma =\tanh \left( \frac{J}{k_{B}T}\right) $, $k_{B}$ being the
Boltzmann constant and $T$ the absolute temperature, and $\alpha _{i}$ is
the inverse of the relaxation time $\tau _{i}$ of spin $i$ in the absence of
spin interactions. It should be pointed out that the rule of transition
stated in Eq. (\ref{2})allows for single site excitations such that the
transitions are not correlated. Further, the choice made in Eq. (\ref{3b})
immediately implies that detailed balance does not hold for this model and
also that not every state in the phase space of the system is accessible. In
particular, one should note that within this model no equilibrium state
exists, although when the time goes to infinity a steady state is eventually
attained. This feature is shared by other purely stochastic models such as
the biased Ising lattice gas and its many variations\cite{Smit and Zia}.

If we now let $\alpha _{1}$ and $\alpha _{2}$ represent the inverses of the
free spin relaxation times of chains composed solely of spins of species $1$
or species $2$, respectively, then we can set $\alpha _{i}=\overline{\alpha }%
_{1}-(-1)^{i}\overline{\alpha _{2}}$ , where $\overline{\alpha }_{1}=\frac{
\alpha _{1}+\alpha _{2}}{2}$ and $\overline{\alpha }_{2}=\frac{\alpha
_{1}-\alpha _{2}}{2}$.

The time dependent probability $P(\sigma _1,\sigma _2,...\sigma _N;t)\equiv
P\left( \left\{ \sigma ^N\right\} ,t\right) $ for a given spin configuration
satisfies the master equation

\begin{eqnarray}
\frac{dP\left( \left\{ \sigma ^N\right\} ,t\right) }{dt} &=&-\sum%
\limits_{i=1}^Nw_i\left( \sigma _{i-1},\sigma _i\right) P\left( \left\{
\sigma ^N\right\} ,t\right)  \nonumber \\
&&+\sum\limits_{i=1}^Nw_i\left( \sigma _{i-1},-\sigma _i\right) P\left(
T_i\left\{ \sigma ^N\right\} ,t\right) .  \label{4}
\end{eqnarray}

The dynamical properties we are interested in require the knowledge of some
moments of the probability $P\left( \left\{ \sigma ^N\right\} ,t\right) $.
Hence, we introduce the following expectation values and correlation
functions defined as:

\begin{equation}
q_i\left( t\right) =\left\langle \sigma _i\left( t\right) \right\rangle
=\sum\limits_{\left\{ \sigma ^N\right\} }\sigma _iP\left( \left\{ \sigma
^N\right\} ,t\right)  \label{5}
\end{equation}

\begin{equation}
r_{i,j}\left( t\right) =\left\langle \sigma _i\left( t\right) \sigma
_j(t)\right\rangle =\sum\limits_{\left\{ \sigma ^N\right\} }\sigma _i\sigma
_jP\left( \left\{ \sigma ^N\right\} ,t\right)  \label{6}
\end{equation}

\noindent and

\begin{eqnarray}
c_{i,j}\left( t^{\prime },t^{\prime }+t\right) =\Theta\left( t \right)
\left\langle \sigma _{i}\left( t^{\prime }\right) \sigma _{j}\left(
t^{\prime }+t\right) \right\rangle =  \nonumber \\
\sum\limits_{\left\{ \sigma ^{N}\right\} ,\left\{ \sigma ^{N\prime }\right\}
}\sigma _{i}^{\prime }P\left( \left\{ \sigma ^{N\prime }\right\} ,t^{\prime
}\right) \sigma _{j}p\left( \left\{ \sigma ^{N}\right\} |\left\{ \sigma
^{N\prime }\right\} ,t\right)  \label{7}
\end{eqnarray}

\noindent where $\Theta\left( t \right)$is the Heaviside step function and
the sums run over all possible configurations compatible with our rule of
motion. The second equality of Eq. (\ref{7}), which gives the formal
definition of the time-delayed correlation function, involves $p\left(
\left\{ \sigma ^{N}\right\} |\left\{ \sigma ^{N\prime }\right\} ,t\right) $,
the conditional probability of the chain having the configuration $\left\{
\sigma ^{N}\right\} $ at time $t^{\prime }+t$ provided it had the
configuration $\left\{ \sigma ^{N\prime }\right\} =\left\{ \sigma
_{1}^{\prime },\sigma _{2}^{\prime },...,\sigma _{N}^{\prime }\right\} $ at
time $t^{\prime }$. Multiplying the master equation by the appropriate
quantities and performing the required summations we obtain the set of time
evolution equations that will be used in our later development. These are
given by

\begin{equation}
\frac{dq_j}{dt}=-2\alpha _j\left( q_j-\gamma q_{j-1}\right)  \label{8}
\end{equation}

\noindent and

\begin{equation}
\frac{dc_{i,j}\left( t^{\prime },t^{\prime }+t\right) }{dt}
=r_{i,j}(t^{\prime })\delta (t)- 2\alpha _{j}\left( c_{i,j}\left( t^{\prime
},t^{\prime }+t\right)- \gamma c_{i,j-1}\left( t^{\prime },t^{\prime
}+t\right) \right) .  \label{9}
\end{equation}

We now impose translational invariance and introduce $\widetilde{q}_k$, the
(spatial) Fourier transform of $q_j$, the $t^{\prime }\rightarrow \infty $
limit of the (temporal) Fourier transform of $c_l\left( t^{\prime
},t^{\prime }+t\right) \equiv c_{i,j}\left( t^{\prime },t^{\prime }+t\right) 
$ (with $l=j-i$) denoted by $\widehat{c}_l(\omega )$, and $\widetilde{C}%
_k(\omega )$, the spatial Fourier transform of $\widehat{c}_l(\omega )$,
defined through

\begin{equation}
q_j=\frac 1{\sqrt{N}}\sum\limits_k\widetilde{q}_k\exp \left( ikj\right) ,
\label{10a}
\end{equation}

\begin{equation}
\widehat{c}_l(\omega )=\lim_{t^{\prime }\rightarrow \infty }\frac 1{2\pi }%
\int_{-\infty }^\infty c_l\left( t^{\prime },t^{\prime }+t\right) \exp
(-i\omega t)dt  \label{11}
\end{equation}

\noindent and

\begin{equation}
\widetilde{C}_k(\omega )\equiv \left\langle \sigma _{-k}\sigma
_k\right\rangle _\omega =\frac 1N\sum_l\widehat{c}_l(\omega )\exp (-ikl).
\label{12}
\end{equation}

In terms of these quantities, Eqs. (\ref{8}) and (\ref{9}) may be rewritten,
respectively, as

\begin{equation}
\frac{d\Psi _k}{dt}={\bf M}_k\Psi _k  \label{13}
\end{equation}
\noindent and

\begin{equation}
i\omega \widehat{c}_l(\omega )=r_l^\infty -2\alpha _l\left( \widehat{c}%
_l(\omega )-\gamma \widehat{c}_{l-1}(\omega )\right) ,  \label{14}
\end{equation}

\noindent where

\begin{equation}
\Psi _k=\left( 
\begin{array}{c}
\widetilde{q}_k \\ 
\widetilde{q}_{k-\pi }
\end{array}
\right) ,  \label{15}
\end{equation}

\begin{equation}
{\bf M}_{k}=\left( 
\begin{array}{cc}
-2\overline{\alpha }_{1}\left( 1-\gamma e^{-ik}\right) & 2\overline{\alpha }%
_{2}\left( 1+\gamma e^{-ik}\right) \\ 
2\overline{\alpha }_{2}\left( 1-\gamma e^{-ik}\right) & -2\overline{\alpha }%
_{1}\left( 1+\gamma e^{-ik}\right)
\end{array}
\right) ,  \label{16}
\end{equation}

\noindent $r_{l}^{\infty }={\lim }_{t\rightarrow \infty }r_{l}(t)$ is the
value of the pair correlation function corresponding to the stationary
solution of the equations of motion in the limit $t\rightarrow \infty$.

The solution to Eq. (\ref{13}), which yields the magnetization, is
straightforward, namely

\begin{equation}
\Psi _k\left( t\right) =e^{{\bf M}_kt}\Psi _k\left( 0\right) .  \label{17}
\end{equation}

The relaxation process of the wave-vector dependent magnetization is
determined by the eigenvalues of ${\bf M}_k$ . These are given by

\begin{equation}
\lambda _{k}^{\pm }=-2\overline{\alpha }_{1}\pm 2\sqrt{\overline{\alpha }
_{1}^{2}-\left( \overline{\alpha }_{1}^{2}-\overline{\alpha }_{2}^{2}\right)
\left( 1-\gamma ^{2}e^{2ik}\right) }.  \label{18}
\end{equation}

It should be stressed that the eigenvalues $\lambda _{k}^{\pm }$ contain
both real and imaginary components. Therefore, the relaxation related to the
real part of the eigenvalues will in general be modulated by the imaginary
component, and care should be taken in defining an adequate correlation
length for the critical dynamics. In fact, one can also associate{\bf \ }%
this correlation length to the modulation of the oscillations. Such
correlation length diverges at the critical point and hence the modulation
eventually dissapears. This kind of behavior has already been thoroughly
discussed in the context of the one-dimensional isotropic ferromagnetic XY
model in an inhomogeneous transverse field \cite{Lind2}. In any case, the
inverses of the ($k$-dependent ) relaxation times $\tau _{k}^{\pm }$ of the $%
\pm k^{th}$ modes are obtained from the real part of $\lambda _{k}^{\pm }$.
In the critical region, that is when $T\rightarrow 0$ and $k\rightarrow 0$, $%
\lambda _{k}^{-}\rightarrow -4\overline{\alpha }_{1}$ while $\lambda
_{k}^{+}\rightarrow 0.$ This means that the critical mode is the one
corresponding to $\lambda _{k}^{+}$. As for the relaxation time, in this
limit one gets

\begin{equation}
\mathop{\rm Re}%
\left( -\lambda _{k}^{+}\right) =-\frac{1}{\tau _{k}}\sim \frac{4\left( 
\overline{\alpha }_{1}^{2}-\overline{\alpha }_{2}^{2}\right) }{\overline{%
\alpha }_{1}}\xi ^{-2}\left[ 1+\frac{(\xi k)^{2}}{2}\right] ,  \label{19}
\end{equation}

\noindent where we have identified the correlation length $\xi $ as $\xi
\sim e^{\frac{J}{k_{b}T}}$ by comparing the former expression with the one
of the dynamic scaling hypothesis $\frac{1}{\tau _{k}}\sim \xi ^{-z}f(\xi k)$
. Therefore we find $z=2$ , which is precisely the same result as for the
alternating isotopic Glauber chain \cite{Lind1}. Note, however, that this
correlation length corresponds to the one of an Ising model with an
effective exchange constant $\frac{J}{2}$ . This can be easily seen by
noting that the equations of motion for the two-spin correlations are in our
case formally identical to those in the Glauber chain, but in the latter $%
\gamma _{G}=\tanh \left( \frac{2J}{k_{B}T}\right) $. Interestingly enough $%
\xi $ also corresponds, as expected, to the correlation length of the steady
state attained by the system. The role of the effective constant $\frac{J}{2}
$ had been already pointed out in the case of the model introduced in ref. 
\cite{LHTGE}.

Now we turn to the calculation of the other interesting response function,
namely the frequency and wave-vector dependent susceptibility $S_{k}\left(
\omega \right) $, which, by virtue of the fluctuation-dissipation theorem 
\cite{Kubo}, is defined by 
\begin{equation}
S_{k}\left( \omega \right) =\frac{\left\langle \sigma _{k}\sigma
_{-k}\right\rangle _{\infty }}{k_{B}T}-\frac{i\omega \left\langle \sigma
_{k}\sigma _{-k}\right\rangle _{\omega }}{k_{B}T},  \label{20}
\end{equation}
where $\left\langle \sigma _{k}\sigma _{-k}\right\rangle _{\infty }=1/\left(
1-\gamma \cos k\right) \cosh \frac{J}{k_{B}T}$ is the static correlation
function and $\left\langle \sigma _{k}\sigma _{-k}\right\rangle _{\omega }$
the Fourier transform of the dynamic one. \noindent After some rather
lengthy but not too complicated algebraic manipulations starting with Eq.(%
\ref{14}) we arrive at

\begin{eqnarray}
&&S_{k}\left( \omega \right) =\frac{1}{k_{B}T\left( 1-\gamma \cos k\right)
\cosh \frac{J}{k_{B}T}}  \nonumber \\
&\times& \left[ 1-\frac{i\omega \left( i\omega +2\overline{\alpha }%
_{1}(1+\gamma e^{-ik})\right) }{\left( i\omega +2\overline{\alpha }%
_{1}\right) ^{2}-4\overline{\alpha }_{2}^{2}+4\gamma ^{2}\left( \overline{%
\alpha }_{2}^{2}-\overline{\alpha }_{1}^{2}\right) e^{-2ik}}\right] .
\label{21}
\end{eqnarray}

The result embodied in Eq. (\ref{21}), constitutes the proper framework in
which to discuss the issue of Nagel scaling in the relaxation of a linear
chain with translational invariance and restricted dynamics. This will be
postponed until section 4. For the time being, we just quote the equivalent
result for the isotopic chain with Glauber dynamics\cite{Lind1},\cite{PRL},
namely

\begin{eqnarray}
&&S_{k}^{G}\left( \omega \right) =\frac{1}{k_{B}T\left( 1-\gamma _{G}\cos
k\right) \cosh \frac{2J}{k_{B}T}}  \nonumber \\
&\times & \left[ 1-\frac{i\omega \left( i\omega +\overline{\alpha }%
_{1}(1+\gamma _{G}\cos k)\right) }{\left( i\omega +\overline{\alpha }%
_{1}\right) ^{2}-\frac{1}{2}\gamma _{G}^{2}\alpha _{1}\alpha _{2}\left(
1+\cos 2k\right) -\overline{\alpha }_{2}^{2}}\right] ,  \label{21b}
\end{eqnarray}
which is also required for such a discussion. In the next section we will
consider the other simple alternating Ising chain including the dynamics
allowed by our rule of transition.

\section{The alternating-bond chain.}

The model consists of a closed linear chain with N sites and it is
characterized by different coupling constants $J_{j}$ (to be specified
below). N is taken to be an even integer and periodic conditions are also
imposed. This model can be applied in the description of dimerized
structures and the Hamiltonian is again of the Ising type, {\it i.e.}

\begin{equation}
H=-\sum\limits_{j=1}^NJ_j\sigma _j\sigma _{j+1}  \label{22}
\end{equation}

\noindent where the{\bf \ }parameters and variables are defined as in the
previous section. Once more and for the same reasons as above, we specify
the Hamiltonian to have a precise definition of the states that intervene in
allowed transitions, but it will play no further relevant role in the
calculations that follow.

Considering again the same kind of biased random walk for the domain wall
motion and the rule of transition given by Eq. (\ref{2}), in this case the
transition probabilities are taken to be

\begin{equation}
w_{i}(\sigma _{i-1},\sigma _{i})=\alpha (1-\beta _{i-1}\sigma _{i-1}\sigma
_{i})  \label{23b}
\end{equation}

\noindent where $\alpha $ is the inverse of the free spin relaxation time, $%
\beta _j=\tanh \left( \frac{J_j}{k_BT}\right) $and $J_j=\frac 12\left(
J_1+J_2\right) -\frac{\left( -1\right) ^j}2\left( J_1-J_2\right) $. Here $%
J_1 $ and $J_2$ represent the coupling constants of two different uniform
Ising chains, respectively. It is also convenient to introduce the
quantities $\gamma _1=\frac 12\left[ \tanh \left( \frac{J_1}{k_BT}\right)
+\tanh \left( \frac{J_2}{k_BT}\right) \right] $ and $\gamma _2=\frac 12\left[
\tanh \left( \frac{J_1}{k_BT}\right) -\tanh \left( \frac{J_2}{k_BT} \right) %
\right] $ so that $\beta _j$ may be expressed as $\beta _j=\gamma _1-\left(
-1\right) ^j\gamma _2$. The analogous forms of Eqs. (\ref{8}) and ( \ref{9}%
), obtained using a similar procedure, read

\begin{equation}
\frac{dq_j}{dt}=-2\alpha \left( q_j-\beta _{j-1}q_{j-1}\right)  \label{24}
\end{equation}

\noindent and

\begin{equation}
\frac{dc_{i,j}\left( t^{\prime },t^{\prime }+t\right) }{dt}
=r_{i,j}(t^{\prime })\delta (t)- 2\alpha \left( c_{i,j}\left( t^{\prime
},t^{\prime }+t\right) -\beta _{j-1}c_{i,j-1}\left( t^{\prime },t^{\prime
}+t\right) \right) .  \label{25}
\end{equation}

As in the previous case, we now impose translational invariance and again
use $\widetilde{q}_{k}$, $\widehat{c}_{l}(\omega )$, and $\widetilde{C}
_{k}(\omega )$ defined as in section 2 to rewrite Eqs. (\ref{24}) and (\ref
{25}) as

\begin{equation}
\frac{d\Phi _k}{dt}={\bf N}_k\Phi _k  \label{26}
\end{equation}
\noindent and

\begin{equation}
i\omega \widehat{c}_l(\omega )=r_l^\infty -2\alpha \left( \widehat{c}
_l(\omega )-\beta _{l-1}\widehat{c}_{l-1}(\omega )\right) ,  \label{27}
\end{equation}

\noindent where

\begin{equation}
\Phi _k=\left( 
\begin{array}{c}
\widetilde{q}_k \\ 
\widetilde{q}_{k-\pi }
\end{array}
\right) ,  \label{28}
\end{equation}

\begin{equation}
{\bf N}_k=\left( 
\begin{array}{cc}
-2\alpha \left( 1-\gamma _1e^{-ik}\right) & -2\alpha \gamma _2e^{-ik} \\ 
2\alpha \gamma _2e^{-ik} & -2\alpha \left( 1+\gamma _1e^{-ik}\right)
\end{array}
\right) .  \label{29}
\end{equation}

The solution to Eq. (\ref{26}) , which yields the magnetization, is again
straightforward, namely

\begin{equation}
\Phi _k\left( t\right) =e^{{\bf N}_kt}\Phi _k\left( 0\right) .  \label{30}
\end{equation}

The eigenvalues of ${\bf N}_{k}$ may be obtained very easily with the result

\begin{equation}
\overline{\lambda }_{k}^{\pm }=-2\alpha \left( 1\pm \sqrt{\gamma
_{1}^{2}e^{-2ik}-\gamma _{2}^{2}e^{-2ik}}\right) ,  \label{31}
\end{equation}

\noindent and again contain real and imaginary components. Proceeding as in
the derivation of Eq.(\ref{19}), we find that the critical mode corresponds
to $\overline{\lambda }_{k}^{-}$ and that the relaxation time is given by

\begin{equation}
\mathop{\rm Re}%
\left( -\overline{\lambda }_{k}^{-}\right) =\frac{1}{\overline{\tau }_{k}}%
\sim 2\alpha \overline{\xi }^{-2}\left[ 1+\frac{(\overline{\xi }k)^{2}}{2}%
\right] .  \label{32}
\end{equation}

Here, considering that $J_{1}>J_{2}$, the correlation length $\overline{\xi }
$ has been identified as $\overline{\xi }\sim e^{\frac{J_{2}}{k_{B}T}}$ by
again comparing Eq. (\ref{32}) with the dynamic scaling hypothesis $\frac{1}{%
\overline{\tau }_{k}}\sim \overline{\xi }^{-z}f(\overline{\xi }k).$
Therefore, we also find $z=2$ in this case, which coincides with the result
derived above for the isotopic chain and, consequently, the restricted
dynamics produces a dynamic critical exponent which is independent of the
kind of interactions in these models. Note that once more the correlation
length $\overline{\xi }$ corresponds to the one of an Ising model with an
effective constant $\frac{J_{2}}{2}$.

The dynamic scaling exponent $z$\ may also be derived in this case by an
alternative argument based on domain wall motion. Consider the limit $%
T\rightarrow 0$\ in which critical slowing-down occurs and domains are
formed. According to Eq. (\ref{23b}), if spin $j$\ is in the interior of the
domain, in this limit $w_{j}(\sigma _{j-1},\sigma _{j})\rightarrow 0$\
irrespective of the value of $\beta _{j-1};$on the other hand, if spin $i$\
belongs to a domain wall, $w_{i}(\sigma _{i-1},\sigma _{i})\rightarrow
2\alpha $\ also independently of $\beta _{i-1}$. Therefore in the low
temperature limit $w_{j}<<w_{i}$, so that the dynamics of the system is
equivalent to a biased random motion of the domain wall for the Hamiltonian
system, in this instance the Ising model. Hence, within restricted dynamics,
the relaxation time $\overline{\tau }_{0}$\ for the decay of a domain of
size $\overline{\xi }$ \ (which is the correlation length associated to the
steady state of the stochastic model) is such that $\overline{\tau }_{0}\sim 
\overline{\xi }^{z}$\ and may be related to a biased random walk \cite
{Cordery}. To this end, let $N_{0}$\ be the number of spins that must flip
to obtain a domain of size $\xi _{I}$\ (the correlation length of the Ising
model). Since the domain has the size of the biased random walk with $N_{0}$%
\ steps and $z=1$\ for the biased random walk, $\xi _{I}\sim N_{0}$. On the
other hand, $\overline{\tau }_{0}\sim N_{0}w_{i}^{-1}\sim \xi _{I}/(2\alpha
) $. But $\xi _{I}=\overline{\xi }^{2}$, so that $z=2$\ for the
alternating-bond model for any choice of $J_{1}$\ and $J_{2}$.

Finally, for the sake of completeness we will also compute the frequency and
wave-vector dependent susceptibility $\overline{\chi }_{k}\left( \omega
\right) $ of this model. Taking the same steps as in the case of the
isotopic chain, we find

\begin{eqnarray}
\overline{S}_{k}\left( \omega \right) &=&\frac{\left\langle \sigma
_{k}\sigma _{-k}\right\rangle _{\infty }}{k_{B}T}\left[ 1-\frac{i\omega
\left( i\omega +2\alpha +2\alpha \gamma _{1}e^{-ik}\right) }{\left( i\omega
+2\alpha \right) ^{2}+4\alpha ^{2}e^{-2ik}\left( \gamma _{2}^{2}-\gamma
_{1}^{2}\right) }\right]  \nonumber \\
&+&\frac{\left\langle \sigma _{k}\sigma _{-k+\pi }\right\rangle _{\infty }}{%
k_{B}T}\frac{2i\omega \alpha \gamma _{2}e^{-ik}}{\left( i\omega +2\alpha
\right) ^{2}+4\alpha ^{2}e^{-2ik}\left( \gamma _{2}^{2}-\gamma
_{1}^{2}\right) }  \label{33}
\end{eqnarray}

\noindent with

\begin{equation}
\left\langle \sigma _{k}\sigma _{-k}\right\rangle _{\infty }=\frac{\left(
1-u_{1}u_{2}\right) \left[ 1+u_{1}u_{2}+\left( u_{1}+u_{2}\right) \cos k%
\right] }{1+u_{1}^{2}u_{2}^{2}-2u_{1}u_{2}\cos 2k},  \label{34}
\end{equation}

\begin{equation}
\left\langle \sigma _{k}\sigma _{-k+\pi }\right\rangle _{\infty }=\frac{%
i\left( 1+u_{1}u_{2}\right) \left( u_{2}-u_{1}\right) \sin k}{%
1+u_{1}^{2}u_{2}^{2}-2u_{1}u_{2}\cos 2k}  \label{35}
\end{equation}
and $u_{j}=\tanh \left( \frac{J_{j}}{2k_{B}T}\right) $, $j=1,2$. Again for
the sake of comparison and in order to correct a misprint in the formula
that appeared in ref. \cite{Lind1}, we quote the equivalent results using
the Glauber dynamics, namely

\begin{eqnarray}
\overline{S}_{k}^{G}\left( \omega \right) &=&\frac{\left\langle \sigma
_{k}\sigma _{-k}\right\rangle _{\infty }^{G}}{k_{B}T}\left[ 1-\frac{i\omega
\left( i\omega +\alpha +2\alpha \gamma ^{^{\prime }G}\cos k\right) }{\left(
i\omega +\alpha \right) ^{2}-2\alpha ^{2}((\gamma ^{^{\prime
}G})^{2}-(\delta ^{G})^{2})\cos 2k-2\alpha ^{2}((\gamma ^{^{\prime
}G})^{2}+(\delta ^{G})^{2})}\right]  \nonumber \\
&+&\frac{\left\langle \sigma _{k}\sigma _{-k+\pi }\right\rangle _{\infty
}^{G}}{k_{B}T}\frac{2\alpha \delta ^{G}\omega \sin k}{\left( i\omega +\alpha
\right) ^{2}-2\alpha ^{2}((\gamma ^{^{\prime }G})^{2}-(\delta ^{G})^{2})\cos
2k-2\alpha ^{2}((\gamma ^{^{\prime }G})^{2}+(\delta ^{G})^{2})}
\end{eqnarray}

\noindent with

\begin{equation}
\left\langle \sigma _{k}\sigma _{-k}\right\rangle _{\infty }^{G}=\frac{%
\left( 1-u_{1}^{G}u_{2}^{G}\right) \left[ 1+u_{1}^{G}u_{2}^{G}+\left(
u_{1}^{G}+u_{2}^{G}\right) \cos k\right] }{%
1+(u_{1}^{G}u_{2}^{G})^{2}-2u_{1}^{G}u_{2}^{G}\cos 2k},
\end{equation}

\begin{equation}
\left\langle \sigma _{k}\sigma _{-k+\pi }\right\rangle _{\infty }^{G}=\frac{%
i\left( 1+u_{1}^{G}u_{2}^{G}\right) \left( u_{2}^{G}-u_{1}^{G}\right) \sin k%
}{1+(u_{1}^{G}u_{2}^{G})^{2}-2u_{1}^{G}u_{2}^{G}\cos 2k},
\end{equation}
$\gamma ^{^{\prime }G}=\frac{1}{2}\tanh \left( \frac{J_{1}+J_{2}}{k_{B}T}%
\right) $, $\delta ^{G}=-\frac{1}{2}\tanh \left( \frac{J_{1}-J_{2}}{k_{B}T}%
\right) $ and $u_{j}^{G}=\tanh \left( \frac{J_{j}}{k_{B}T}\right) $, $j=1,2$%
. This concludes our analysis of the alternating-bond chain with the
restricted dynamics.

\section{Susceptibility and scaling behavior.}

Thus far, we have examined the behavior of the critical dynamic exponent. In
this section we will continue our exploration by considering the scaling
properties of the susceptibility in the isotopic alternating chain. This is
most conveniently done through the quantity $\chi (\omega )\equiv
k_{B}TS_{0}(\omega )/\left\langle \sigma _{0}\sigma _{0}\right\rangle
_{\infty }$ . It should be noted that if we look at the uniform chain, {\it %
i.e. }we set $\alpha _{1}=\alpha _{2}$ in Eq. (\ref{21}) and take the limit $%
k=0$ in the resulting expression, $\chi $ obeys the usual normalized Debye
scaling : if one scales the frequency with the inverse of the (single)
relaxation time and the real and imaginary parts are then divided by their
values at zero and one, respectively, one gets a universal curve. However,
as mentioned in the Introduction, recently experimental work on dielectric
relaxation has been reported in terms of a new scaling function \cite{Nagel}%
- \cite{Wu} which is thought to be related to multifractal scaling. In this
new scaling, the abscissa is $(1+W)\log _{10}\left( \omega /\omega
_{p}\right) /W^{2}$ and the ordinate is $\log _{10}\left( \chi ^{\prime
\prime }(\omega )\omega _{p}/\omega \Delta \chi \right) /W$. Here, $\chi
^{\prime \prime }$ is the imaginary part of $\chi (\omega )$, $W$ is the
full width at half maximum of $\chi ^{\prime \prime }$, $\omega $ is the
frequency and $\omega _{p}$ the one corresponding to the peak in $\chi
^{\prime \prime }$, and $\Delta \chi =\chi (0)-\chi _{\infty }$ is the
static susceptibility. It is therefore interesting to see whether the
isotopic alternating chain with the restricted dynamics leads to Nagel
scaling \ in the same way that the model with Glauber dynamics does\cite{PRL}%
.

Using Eqs. (\ref{18}) and (\ref{21}) with $k=0$, $\chi $ can be expressed in
the form

\begin{equation}
\chi (\omega )=\frac{(1-\gamma )(\alpha _{1}+\alpha _{2})}{2}\left[ \frac{%
1-f(\alpha _{1},\alpha _{2},\gamma )}{i\omega -\lambda _{0}^{+}}-\frac{%
1+f(\alpha _{1},\alpha _{2},\gamma )}{i\omega -\lambda _{0}^{-}}\right] .
\label{22m}
\end{equation}
Here, the (temperature dependent) function $f(\alpha _{1},\alpha _{2},\gamma
)$ is given by

\begin{equation}
f(\alpha _{1},\alpha _{2},\gamma )=\frac{(\alpha _{1}-\alpha
_{2})^{2}-4\alpha _{1}\alpha _{2}\gamma ^{2}}{(\alpha _{1}+\alpha _{2})\sqrt{%
(\alpha _{1}-\alpha _{2})^{2}+4\alpha _{1}\alpha _{2}\gamma ^{2}}}.
\label{23m}
\end{equation}
It should be pointed out that for the case $\gamma =0,$ we get

\begin{equation}
\chi _{\gamma =0}=\frac{\alpha _{1}}{i\omega +2\alpha _{1}}+\frac{\alpha _{2}%
}{i\omega +2\alpha _{2}},  \label{24m}
\end{equation}
so that the general structure of the result for the susceptibility of the
alternating isotopic chain is preserved irrespective of the value of $\gamma 
$ ({\it i.e.} of the temperature), namely a linear combination of two
Debye-like terms.

With the aid of Eqs. (\ref{22m}) and (\ref{23m}) in Figs. 1 to 4 we present
Nagel plots for the cases $\alpha _{1}=\alpha _{2}=1$, $\alpha _{1}=1$ and $%
\alpha _{2}=2$, $\alpha _{1}=1$ and $\alpha _{2}=100$, and $\alpha _{1}=1$
and $\alpha _{2}=1000$, respectively, and different values of $1/T^{\ast
}\equiv J/k_{B}T$. We also include in these figures plots of $\chi ^{\prime
\prime }(\omega )$ vs. $\omega /\omega _{p}$ which are the natural variables
of the Debye relaxation. While the first case (which as stated above
corresponds to Debye behavior) does not show Nagel scaling, the situation
somewhat improves in the second one (where clearly improvement means less
dispersion in the curves), and when the two relaxation times are not only
different but very far apart (third and fourth cases) the scaling is
virtually perfect, provided the temperature lies above some certain critical
value. For comparison, in Fig. 5 we show parallel results computed for the
alternating isotopic chain with Glauber dynamics ({\it c.f. }Eq. (\ref{21b}%
)), for the case $\alpha _{1}=1$ and $\alpha _{2}=100$. The similarity of
the results of both models provides support to the idea that, irrespective
of the specific dynamics, the coexistence of different relaxation mechanisms
lies behind the Nagel scaling and that this only occurs if a threshhold
temperature is surpassed. Nevertheless, the character of the different
dynamics manifests itself in that the slopes of the decaying parts of the
curves in the Nagel plot differ and that, for the same values of $\alpha
_{1} $ and $\alpha _{2}$, the scaling is more closely followed by the model
with the restricted dynamics ({\it c.f. }Figs. 3 and 5).

It should be noted that, as the insets of these figures indicate, both types
of dynamics lead to two peaks in $\chi ^{\prime \prime }(\omega )$ due to
the presence of two different relaxation times.This feature has been also
observed experimentally by Dixon {\it et al}\cite{Nagel}{\it ,}
Lesley-Pelecky and Birge\cite{Birge} and Wu {\it et al}\cite{Wu} in
different materials and associated to the $\alpha $ and $\beta $
relaxations. Very recently it has also been confirmed in experiments by
Brand {\it et al}\cite{Brand} who have also noticed the presence of a third
relaxation process unexplained so far.

\section{Concluding Remarks.}

In this paper we have addressed relaxation processes and the question of the
value of the dynamical critical exponent in kinetic Ising models on
alternating linear chains. Two different issues have been examined in this
context. In the first one, we have shown that with the restricted dynamics
implied by the transition probabilities ({\it c.f.} Eqs. (\ref{3b}) and (\ref
{23b})), both in the alternating-bond chain and in the isotopic chain the
dynamic critical exponent $z$ turns out to be exactly two. This does not
occur if the Glauber dynamics is employed. Hence, the value of the dynamic
critical exponent and the dynamics implied by the rule of transition in
kinetic models are deeply related. As for the second issue, the analysis of
the Nagel plots in the case of the alternating isotopic chain indicates that
the presence of at least two different relaxation mechanisms is required for
the scaling of the susceptibility. This feature agrees with what one finds
with the usual Glauber dynamics\cite{PRL}, as well as the appearance of
plateau regions in the plot if the relaxation times are widely separated and
of the existence of a critical temperature below which the scaling is not
followed. One may reasonably wonder at this stage whether the
alternating-bond chain with the restricted dynamics also obeys the Nagel
scaling. Since for high temperatures this model is equivalent to a chain
with a single relaxation time, it is not surprising that the scaling is not
followed in this instance. This we have confirmed numerically for a variety
of values for $J_{1\text{ }}$and $J_{2}$.{\tt \ }

Although our results for the isotopic chain suggest that the dynamics seems
not to play a key role for the scaling to hold, one should bear in mind that
the isotopic chain with restricted dynamics containing only two isotopes is
somewhat peculiar, so that no definite conclusions on this issue can be
reached at this stage.{\tt \ }Concerning the differences, it is conceivable
that the stochasticity may well be behind the fact that the slope of the
decaying parts of the curves in the Nagel plots is larger for the restricted
dynamics than for the Glauber dynamics. In this respect, it is important to
point out that the experiments in which the Nagel plots have been more
succesful concern glass-forming systems and \ that our original model was
built so as to reflect the topological constraints that are assumed to be
crucial in such systems. It is clear that the restricted dynamics related to
such constraints is also not enough to get the scaling, as exemplified by
the case of the uniform chain $\alpha _{1}=\alpha _{2}=1$. The true
alternating isotopic chain with restricted dynamics examined here, on the
other hand, includes both ingredients and provides a {\it bona fide
microscopic model} in which the Nagel scaling is shown to arise. Moreover,
the fact that we get a bigger slope in the restricted model is consistent
with the experimental finding that such a slope is bigger for the
orientationally disordered crystalline phase of cyclo-octanol ({\it c.f. }%
Lesley-Pelecky and Birge\cite{Birge}) than in the originally studied linear
polymers\cite{Wu}.

Notwithstanding the limitations of this model, our expectation is that both
our earlier results \cite{LHTGE},\cite{PRL} as well as the present ones
provide some insight into the physical origin and validity of the hypothesis
concerning the need of the simultaneous presence of multiple relaxation
mechanisms as related to the proposal of the Nagel plots. A future challenge
is to examine whether the Nagel scaling is also present in other Ising
models recently studied in connection with glassy dynamics\cite{Schulz}
which are based on the spin facilitated models originally introduced by
Fredrickson and Andersen \cite{FA}. Finally, one can conjecture that the
appearance of a third relaxation process as observed in the recent
experiments of dielectric relaxation by Brand {\it et al}\cite{Brand} may be
hopefully catered for within our model through the inclusion of a third
relaxation time. The investigation of this conjecture is presently in
progress.


{\acknowledgments
One of us (L. L. Gon\c{c}alves) wants to thank the brazilian agencies CNPq
and FINEP for partial financial support. The work of another of us (J.
Tag\"{u}e\~{n}a-Mart\'{i}nez) has been supported by DGAPA-UNAM under project
IN-104598. Thanks are also due to Prof. R. B. Stinchcombe for enlightening
discussions on the subject matter of this paper.}

\strut

{\bf Figure captions}

Figure 1. Nagel plot for the case of a uniform chain ($\alpha _{1}=\alpha
_{2}=1$). Note that the proposed scaling does not hold in this case while,
as the inset shows, the susceptibility obeys the usual Debye scaling.

Figure 2. Nagel plot for $\alpha _{1}=1$ and $\alpha _{2}=2$. Here one can
see an improvement of the scaling behavior as compared to the uniform chain
case, except at low $T^{\ast }$. In the inset the plot to test the
performance with respect to the Debye scaling is presented.

Figure 3. Same as Fig. 2 but for the choice $\alpha _{1}=1$ and $\alpha
_{2}=100$. Except at low $T^{\ast }$ values, the trend of improvement of the
agreement with the Nagel scaling is apparent, while the opposite happens
with respect to the Debye scaling.

Figure 4. Same as Figs. 2 and 3 but for $\alpha _{1}=1$ and $\alpha
_{2}=1000 $. The scaling is virtually perfect in this case, except at low $%
T^{\ast }$. Notice the explicit appearance of a plateau region in the plot.
The behavior is here definitely non-Debye.

Figure 5. Nagel plot for an alternating isotopic chain with Glauber dynamics
with$\ \alpha _{1}=1$ and $\alpha _{2}=100$ and different values of the
parameter $T_{G}^{\ast }\equiv k_{B}T/2J$. Here, $\chi ^{G}(\omega )\equiv
k_{B}TS_{0}^{G}(\omega )/\left\langle \sigma _{0}\sigma _{0}\right\rangle
_{\infty }^{G}$ \ with $\left\langle \sigma _{0}\sigma _{0}\right\rangle
_{\infty }^{G}=1/\left( 1-\gamma _{G}\right) \cosh \frac{1}{T_{G}^{\ast }}$
and $S_{0}^{G}(\omega )$ computed with the aid of eq. (\ref{21b}). Note the
similarity of these results with respect to those of Fig. 3.


\newpage

\begin{references}
\bibitem[*]{mariano}  Also Consultant at Programa de Simulaci\'{o}n
Molecular del Instituto Mexicano del Petr\'{o}leo.

\bibitem{Hohenberg}  B. I. Halperin and P. C. Hohenberg, Phys. Rev. {\bf 177}%
, 952 (1969); P. C. Hohenberg and B. I. Halperin, Rev. Mod. Phys. {\bf 49},
435 (1977).

\bibitem{Zhu}  An odd case in this respect is the Gaussian model. Very
recently, it has been shown that in this model $z$ has the same value
independent of spatial dimensionality, J. - Y. Zhu and Z. R. Yang, Phys.
Rev. E {\bf 59}, 1551 (1999).

\bibitem{Kimball}  J. C. Kimball, J. Stat. Phys. {\bf 21}, 289 (1979).

\bibitem{Thol}  U. Deker and F. Haake, Z. Phys. B {\bf 35}, 281 (1979); F.
Haake and K. Thol, Z. Phys. B {\bf 40}, 219 (1980).

\bibitem{Cordery}  R. Cordery, S. Sarker and J. Tobochnick, Phys. Rev. B 
{\bf 24}, 5402 (1981).

\bibitem{Menyhárd}  N. Menhy\'{a}rd and G. \'{O}dor, J. Phys. A: Math. Gen. 
{\bf 28}, 4505 (1995).

\bibitem{Lind1}  L. L. Gon\c{c}alves and N. T. de Oliveira, Can. J. Phys. 
{\bf 63}, 1215 (1985).

\bibitem{Droz}  M. Droz, J. Kamphorst Leal da Silva and A. Malaspinas, Phys.
Lett. A {\bf 115}, 448 (1986); M. Droz, J. Kamphorst Leal da Silva, A.
Malaspinas and A. L. Stella, J. Phys. A: Math. Gen. {\bf 20}, L387 (1987).

\bibitem{Robin}  J. A. Ashraff and R. B. Stinchcombe, Phys. Rev. B {\bf 40},
2278 (1989).

\bibitem{Glauber}  R. J. Glauber, J. Math. Phys. {\bf 4}, 294 (1963).

\bibitem{Eduardo1}  E. J. S. Lage, J. Phys. C: Solid State Phys. {\bf 20},
3969 (1987).

\bibitem{Luscombe}  J. H. Luscombe, Phys. Rev. B {\bf 36}, 501(1987).

\bibitem{Auriac}  J. C. Angles d'Auriac and R. Rammal, J. Phys. A: Math.
Gen. {\bf 20}, 763 (1988).

\bibitem{Eduardo2}  J. M. Nunes da Silva and E. J. S. Lage, J. Stat. Phys. 
{\bf 58}, 115 (1990).

\bibitem{Cornell}  S. Cornell, M. Droz and N. Menyh\'{a}rd, J. Phys. A:
Math. Gen. {\bf 24}, L201 (1991).

\bibitem{Kimmo}  S. J. Cornell, K. Kaski and R. B. Stinchcombe, J. Phys. A:
Math. Gen. {\bf 24}, L865 (1991).

\bibitem{Jefferson}  J. Kamphorst Leal da Silva, A. G. Moreira, M.
Silv\'{e}rio Soares and F. C. S\'{a} Barreto, Phys. Rev. E {\bf 52}, 4527
(1995).

\bibitem{Tong}  P. Tong, Phys. Rev. E {\bf 56}, 1371 (1997).

\bibitem{Southern}  B. W. Southern and Y. Achiam, J. Phys. A: Math. Gen. 
{\bf 26}, 2505 (1993); J. Phys. A: Math. Gen. {\bf 26}, 2519 (1993);

\bibitem{Achiam}  Y. Achiam, {\it Phys. Rev. B} {\bf 31}, 4732 (1985).

\bibitem{LHTGE}  M. L\'{o}pez de Haro, J. Tag\"{u}e\~{n}a-Mart\'{i}nez, B.
Espinosa and L. L. Gon\c{c}alves, J. Phys. A: Math. Gen. {\bf 26}, 6697
(1993). {\bf 29}(E), 7353 (1996); M. L\'{o}pez de Haro, L. L. Gon\c{c}alves
and J. Tag\"{u}e\~{n}a-Mart\'{i}nez, Mod. Phys. Lett. {\bf B 10}, 1441
(1996).

\bibitem{GdM}  J. H. Gibbs and E. A. di Marzio,{\it \ J. Chem. Phys. }{\bf 28%
}, 373 (1958).

\bibitem{Nagel}  P. K. Dixon, L. Wu and S. R. Nagel, B. D. Williams and J.
P. Carini, {\it \ Phys. Rev. Lett.} {\bf 65}, 1108 (1990); M. D. Ediger, C.
A. Angell and S. R. Nagel,{\it \ J. Phys. Chem.} {\bf 100}, 13200 (1996).

\bibitem{Birge}  D. L. Lesley-Pelecky and N. O. Birge, {\it Phys. Rev. Lett.}
{\bf 72}, 1232 (1992).

\bibitem{Wu}  L. Wu, P. K. Dixon, S. Nagel, B. D. Williams, and J. P.
Carini, J. Non-Cryst. Solids {\bf 131-133}, 32 (1991).

\bibitem{PRL}  L. L. Gon\c{c}alves, M. L\'{o}pez de Haro, J. Tag\"{u}e\~{n}%
a-Mart\'{i}nez, \ and R. B. Stinchcombe, {\it Phys. Rev. Lett.} {\bf 84},
1507 (2000).

\bibitem{Skinner}  J. L. Skinner, {\it \ J. Chem. Phys}. {\bf 79}, 1955
(1983).

\bibitem{Kaye}  See for instance B. H. Kaye, {\it A Random Walk Through
Fractal Dimensions}, 2nd. edn. (VCH, Weinheim,1994).

\bibitem{Smit and Zia}  B. Schmittman and R. K. P. Zia, {\it Statistical
Mechanics of Driven Diffusive Systems,} vol. 17 of {\it Phase Transitions
and Critical Phenomena,}\  C. Domb and J. L. Lebowitz, eds. (Academic Press,
New York, 1995){\tt .}

\bibitem{Lind2}  J. P. de Lima and L. L. Gon\c{c}alves, {\it Mod. Phys.
Lett. B }{\bf 14}\&{\bf 15}, 871 (1994).

\bibitem{Kubo}  M. Suzuki and R. Kubo, {\it J. Phys. Soc. Jpn}. {\bf 24}, 51
(1968).

\bibitem{Brand}  R. Brand, P. Lunkenheimer, U. Schneider, and A. Loidl,
cond-mat/0002460 (29 Feb 2000)

\bibitem{Schulz}  E. Follana and F. Ritort, Phys. Rev. B {\bf 54}, 930
(1996); M. Schulz and S. Trimper, J. Stat. Phys. {\bf 58}, 173 (1999); B.
Zheng, M. Schulz and S. Trimper, Phys. Rev. B {\bf 54}, 6717 (1999).

\bibitem{FA}  G. H. Fredrickson and H. C. Andersen, Phys. Rev. Lett. {\bf 53}%
, 1244 (1984); J. Chem. Phys. {\bf 84}, 5822 (1985).
\end{references}
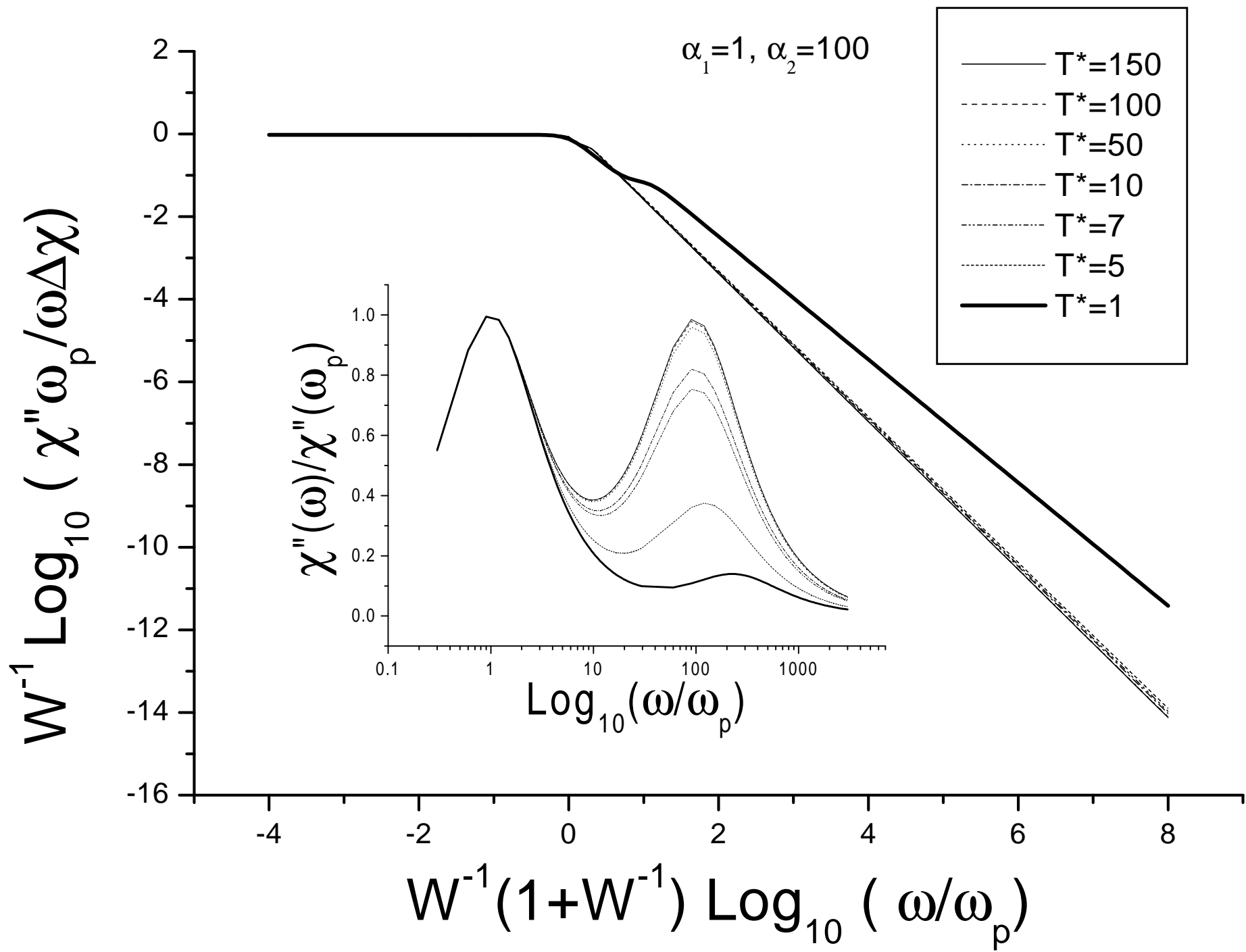
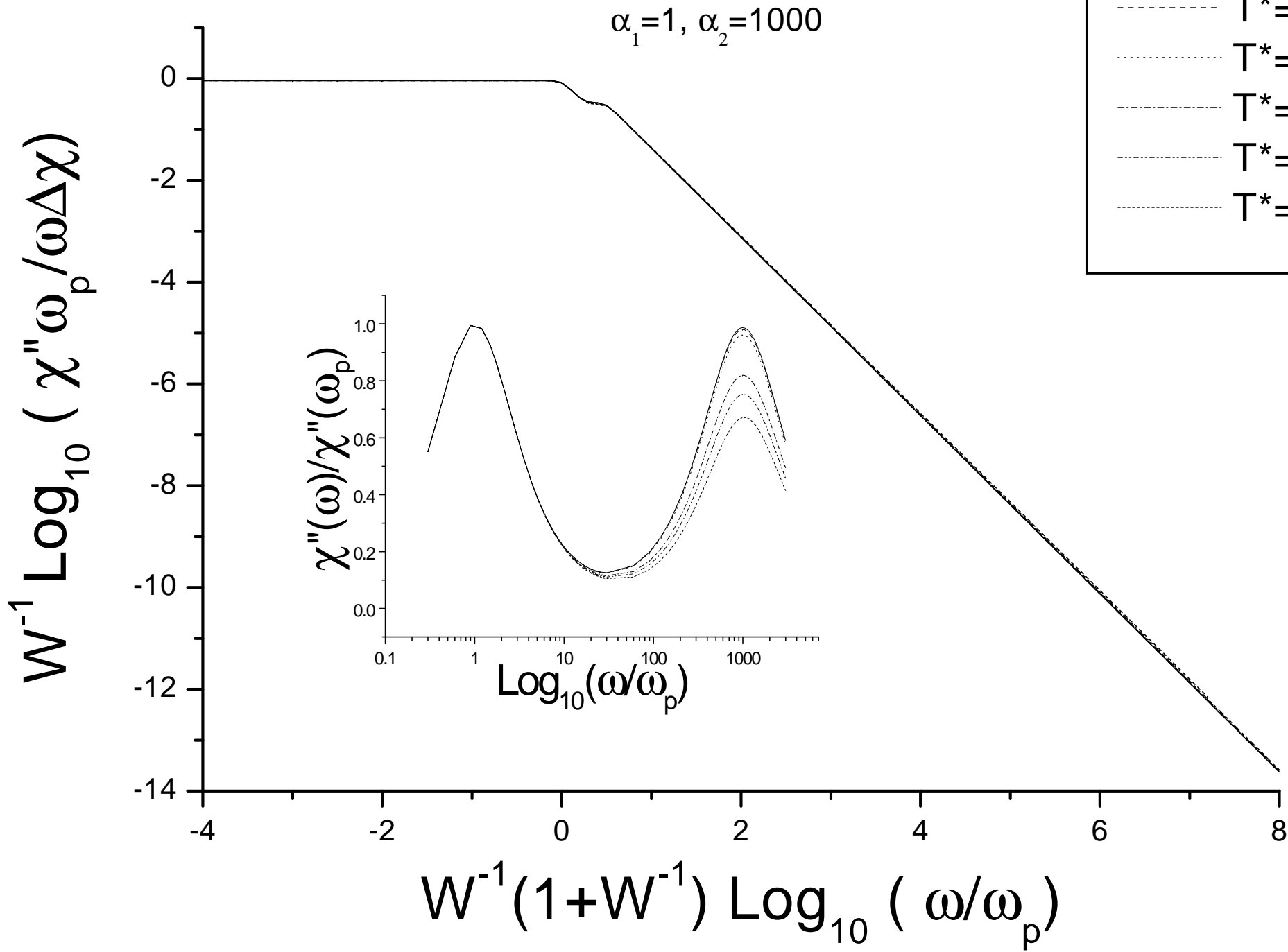
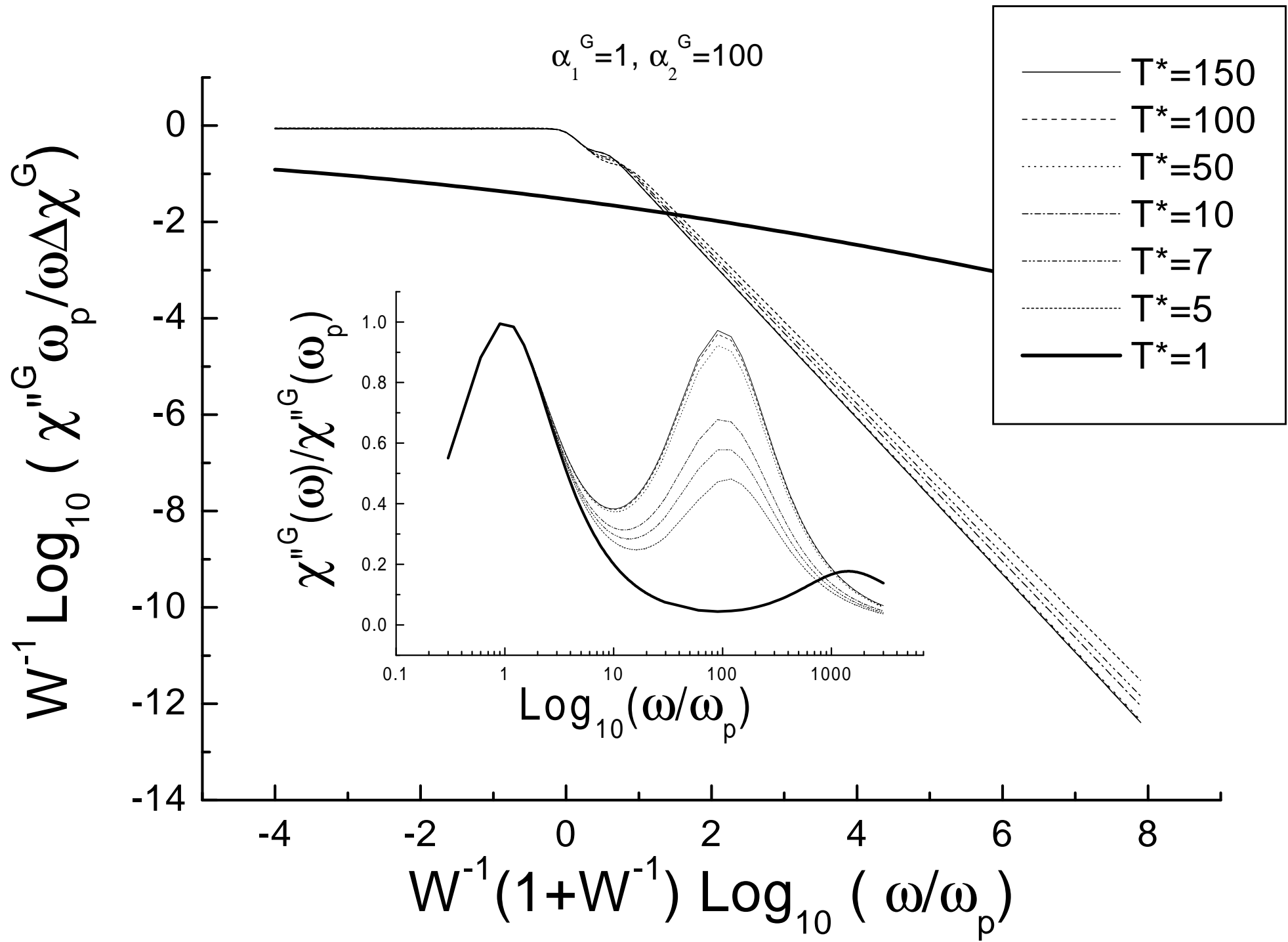
\end{document}